\documentclass[aps,prb,amsfonts,amssymb,twocolumn,amsmath,preprintnumbers,nofootinbib,superscriptaddress,floatfix,showpacs]{revtex4}
\usepackage{amsmath,bm,amsfonts,amssymb}

\begin{document}

\title{Spin-dependent quasiparticle reflection and bound states at
interfaces with itinerant antiferromagnets}

\author{I. V. Bobkova}
\affiliation{Institute of Solid State Physics, Chernogolovka,
Moscow reg., 142432 Russia}
\author{P. J. Hirschfeld}
\affiliation{Department of Physics, University
of Florida, Gainesville, FL 32611 USA}
\author{Yu. S. Barash}
\affiliation{Institute of Solid State Physics, Chernogolovka,
Moscow reg., 142432 Russia}

\date{\today}

\begin{abstract}
We present a formulation of the quasiclassical theory of junctions
between itinerant antiferromagnets (AF) and $s$-wave (sSC) and
$d$-wave superconductors (dSC). For the simplest two-sublattice
antiferromagnet on a bipartite lattice, we derive Andreev-type
equations and show that their solutions lead to a novel channel of
quasiparticle reflection. In particular, quasiparticles in a
normal metal with energies less than or comparable to the
antiferromagnetic gap experience spin-dependent retroreflection at
antiferromagnet-normal metal (AF/N) transparent $(100)$ and
$(110)$ interfaces. A relative phase difference of $\pi$ between
up spin and down spin quasiparticle reflection amplitudes is shown
to lead to zero-energy interface bound states on AF/sSC
interfaces. For an sSC/AF/sSC junction, these bound
states are found to be split, due to a finite width of the AF
interlayer, and carry the supercurrent.  At AF/dSC interfaces we
find no zero-energy bound states for both interface orientations
we considered, in contrast with the case of (110) impenetrable
surface of a dSC.
\end{abstract}
\pacs{74.45.+c, 74.50.+r}
\maketitle
{\it Introduction}. The study of electronic
properties of superconductor-magnetic interfaces has enjoyed a
renaissance in recent years, with increased interest in
technological applications and advances in theory. Ferromagnetic
layers can spin polarize quasiparticle currents and Zeeman split
surface densities of states, with possible applications in
spintronics. SFS junctions have been shown to display $0-\pi$
transitions with varying temperature, width, or orientational
structure of magnetization of the ferromagnetic interlayer
\cite{ryazanov01,aprili02,aprili03,aprili04,
buz82,buz92,nazarov01,bb02,bb01}. The quasiclassical theory of
superconductivity, complemented with appropriate boundary
conditions\cite{mrs88,fog00}, has proven very successful in
dealing with problems like superconducting-ferromagnetic
interfaces. No quasiclassical theory of the antiferromagnetic
state and, in particular, AF/SC interface exists, however.

On the other hand, there are many situations of fundamental and
practical interest which involve antiferromagnet-superconductor
interfaces (AF/SC).  Initial results of experimental
investigations of proximity and Josephson effects through this
type of interfaces have been obtained only recently
\cite{blamire03}. Some unusual properties for specific
antiferromagnetic Josephson weak links have been studied
theoretically for barriers built on a giant magnetoresistance
multilayer or made of doped manganites in the metallic A-phase
\cite{gorkov01}. Many of the properties of high-temperature
superconducting (HTS) cuprate materials  are also thought to
result from a competition between antiferromagnetic and
superconducting order, and there are many naturally occurring
situations and possible devices which might involve such
boundaries.  These include interfaces of insulating and highly
doped cuprates or SC/AF/SC junctions, HTS grain boundaries where
antiferromagnetism may play a role as a surface state, and the
antiferromagnetism which has been observed in HTS vortex cores.

Quasiclassical equations  describe spatial variations of physical
quantities taking place over length scales which are long compared
to atomic distances. The boundary conditions match slowly varying
quantities at interfaces, where the equations themselves do not
apply.  We argue below that the assumption of slow variation of
the sublattice magnetization allows a formulation of the
quasiclassical approach to itinerant antiferromagnets, which we
study with mean field theory. We have found a novel spin-dependent
channel of normal metal (N) quasiparticle reflections from AF/N
interfaces, associated with the structure of the antiferromagnetic
order parameter as well as with a magnetic periodicity along the
interface with  component ${Q}_y$ of the antiferromagnetic wave
vector. Parallel to the interface, components of crystal momenta
for outgoing and incoming quasiparticles differ by ${Q}_y$ in the
novel reflection channel, while differences between components
normal to the interface depend on the shape of the Fermi surface.
Specific channels of quasiparticle reflection have been theoretically
found also in charge density waves (CDW) - normal metal junctions
\cite{kasatkin84,visscher96,rejaei96,ar97}. An excess
quasiparticle reflection below the CDW gap in NbSe$_3$ has
been experimentally identified in Ref.\ \onlinecite{latyshev96}.
We show below, for the simplest cases of (100) and (110)
interfaces, that the spin dependence of quasiparticle reflections
from  boundaries with itinerant antiferromagnets results in
dramatic consequences which do not appear at the nonmagnetic
interfaces mentioned above, however.  An important distinctive
feature of the quasiparticle reflection we study below is the
phase difference $\pi$ of reflection amplitudes taken for spin-up
and spin-down quasiparticles in the normal metal at an AF/N
interface. As a result of this "$\pi$-shift", zero-energy
interface bound states arise on AF/sSC transparent interfaces. By
contrast, a $d$-wave superconductor, which in the case of a free
(110) surface manifests Andreev zero energy bound states, shows no
zero-energy states at AF/dSC transparent interfaces. For an
sSC/AF/sSC junction, the zero-energy states are found to be split
and carry a supercurrent. They can form low-temperature anomalies
in the Josephson critical current through sSC/AF/sSC junctions.
They also give rise to the zero-bias anomaly of the conductance of
N/AF/sSC junctions.

{\it Quasiparticle reflection at a smooth AF interface.}
We begin by considering electrons hopping on a square lattice with
basis vectors $\hat a$ and $\hat b$ (lattice constant $a$), with
superconducting pairing $\Delta^{ij}$ and magnetization $m^i$.
\begin{eqnarray}
H&=&-t\sum_{\langle ij\rangle\sigma} c_{i\sigma}^\dagger
c_{j\sigma} + \sum_{ i,j}
(\Delta^{ij}c_{i\uparrow}^\dagger c_{j\downarrow}^\dagger
+h.c.) \nonumber \\&& + \sum_i m^i (n_{i\uparrow}-n_{i\downarrow})-\mu\sum_i
c_{i\sigma}^\dagger c_{i\sigma}\label{H}
\end{eqnarray}
We assume nearest neighbour hopping, and consider either $s$-wave
pairing $\Delta^{ij}=-V_s\langle
c_{i\downarrow}c_{i\uparrow}\rangle\delta_{ij}=\delta_{ij}\Delta_s^{i}$
or $d$-wave pairing $\Delta^{ij}=-V_d \langle
c_{i\downarrow}c_{j\uparrow}\rangle
=\Delta_d^{ij}\delta_{|i-j|,1}$ such that $\Delta_d^{i i\pm{\hat
a}}=-\Delta_d^{i i\pm{\hat b}}$.  The magnetization
$m^i=-(V_m/2)\langle n_{i\uparrow}-n_{i\downarrow}\rangle$ as a
function of site $i$ is assumed to exhibit  antiferromagnetic
order, such that for the infinite lattice in the absence of local
perturbations, $m^j =(-1)^{j_a+j_b} m$. In the more general case,
the sublattice magnetization $m$ is assumed to be slowly varying
function on scales of the lattice spacing $a$. We assume always
that $m$ is nonzero only on one semi-infinite half-space, while
$\Delta$ may be nonzero on the other; thus there is generally no
additional potential barrier between the two systems, although we
discuss the possibility.

The normal state electron band $\xi({\bf k})=-\mu - 2 t (\cos k_a
+\cos k_b)$ and the respective Brillouin zone is spanned by
$k_{a,b} \in [-\pi,\pi]$, where momenta are given in units of
$a^{-1}$. In describing quasiparticle reflections, it is
convenient also to work in a coordinate system where $x$ and $y$
describe coordinates perpendicular and parallel to the interface,
respectively. For a (100) interface, the band and the zone have
the same forms in the $x,y$-coordinates. For a (110) interface,
however, we have $\xi({\bf k})=-\mu - 4t\cos\bigl(k_x/\sqrt{2}\bigr)
\cos\bigl(k_y/\sqrt{2}\bigr)$ and $k_{x}\in[-\sqrt{2}\pi,\sqrt{2}\pi]$,\,
$k_{y}\in[-\pi/\sqrt{2},\pi/\sqrt{2}]$, on account of
the periodic conditions along the surface.

A specific feature of the antiferromagnetic state, which is
important in the derivation of quasiclassical equations, is the
presence of rapidly oscillating term $m^j =(-1)^{j_a+j_b}
m=\exp(i{\bm Q}{\bm j})m$, which results in
slowly varying Andreev amlitudes with wave vector ${\bm k}+ {\bm
Q}$ in the equations for the amplitudes with the wave vector $\bm
k$. In the case in question,  $2\bm Q$ coincides with a basis
vector of the reciprocal lattice of the nonmagnetic crystal.
For this reason the quasiclassical equations can
be written for pairs of entangled quasiparticle trajectories $\bm
k$ and ${\bm k} +{\bm Q}$. As usual in the quasiclassical theory
of superconductivity, we require that $\xi_s\equiv \hbar v_F/
\Delta_{s,d}\gg a$, and similarly require that the magnetic
``coherence length" $\xi_m\equiv \hbar v_F/|m|\gg a$.  We also
assume that the deviation from
half-filling in the antiferromagnet is not large $\mu\ll
\epsilon_F$. Otherwise the antiferromagnetic state would be
unstable within the framework of a generic Hubbard-like model.
Then $\mu$ should be included directly in the quasiclassical
equations, not in the rapidly oscillating exponentials. Under this
condition, the nesting relation is valid with  quasiclassical
accuracy and energies of normal state quasiparticles with momenta
$\bm k_F$ and ${\bm k_F}+{\bm Q}$ both lie on the Fermi surface.

It is now convenient to  collect into a Nambu 4-spinor the
Andreev amplitudes $\psi^{T}_j\equiv (u_{j\sigma}(\bm k_F), u_{j\sigma}(\bm k_F
+ \bm Q),v_{j{\bar\sigma}}(\bm k_F),v_{j{\bar\sigma}}(\bm k_F+
\bm Q))$, and to define the Pauli matrices $\rho_\alpha$, $\tau_\alpha$
in space of two quasiparticle trajectories ($\bm k$, $\bm k + \bm Q$) and
in particle-hole space respectively.

Then the Andreev equations take the form:
\begin{eqnarray}
\left(- \mu \tau_3 \rho_0 -i\tau_3\rho_3 v_{F,x}\dfrac{\partial}{\partial
x} + \sigma{m}(x)\tau_0\rho_1 +{\check \Delta}(x)\right) {\psi}(x)=\nonumber \\
= \varepsilon {\psi}(x).
\label{andr}
\end{eqnarray}
Here $v_{F,x}=\frac{\partial \xi({\bm k}_F)}{\partial k_x}\bigl|_{\mu=0}$ is
the Fermi velocity at half-filling, $\check
\Delta(x)=\check \Delta_s(x) +\check \Delta_d({\bm k}_F,x)$,
$\check\Delta_s(x)=\sigma\rho_0\Delta_s(x)
\dfrac{\tau_+}{2}+\sigma\rho_0\Delta^{*}_s(x)\dfrac{\tau_-}{2}$,
$\check\Delta_d({\bm k}_F,x)=\sigma\Delta_d({\bm k}_F,x)\rho_3
\dfrac{\tau_+}{2}+\sigma\Delta^{*}_d({\bm k}_F,x)\rho_3
\dfrac{\tau_-}{2}$. Quasiclassical Eqs.(\ref{andr}) do not apply
in vicinity of quasiparticle momenta where $v_{F,x}=0$. In
particular, they do not apply near saddle points of quasiparticle
energies where Van Hove singularities of the normal metal density
of states take place. Since we will be interested mostly in
transport across the interface, where the additional factor
$v_{F,x}$ arises, these momenta do not contribute to the results
noticeably and the conditions turn out not to be restrictive.
Eqs.(\ref{andr}) are written for a quasiparticle spin direction
$\sigma$ ($\sigma=-\bar\sigma=\pm1$), with the quantization axis
along the magnetization at site $j_a=j_b=0$. The quasiclassical
equations for the superconducting side can be formulated also for
$\mu\sim\varepsilon_F$. The parameter $\mu$ does not enter
quasiclassical equations in this case and leads to a large Fermi
velocity mismatch on the boundary. We have presented only the
result (\ref{andr}) valid for $\mu\ll \varepsilon_F$, since we
discuss primarily the case of a small Fermi velocity mismatch
between a superconductor and an antiferromagnet. Eq.(\ref{andr})
also implies that no subdominant pairing channels are important in
the problem, which we solve below with step-like
(nonselfconsistent) profiles of the order parameters.

{\it Normal/AF interface.} We now apply Eqs.(\ref{andr}) to the
problem of electron reflection at AF/N interfaces. The initial and
final states of an electron reflected from AF/N interface belong
to the bulk of the normal metal. We find that the incoming
electron with a momentum ${\bm k}_F$ acquires in a reflection
event the momentum ${\bm k}_F+{\bm Q}$. The antiferromagnetic wave
vector on the square lattice is ${\bm Q}=(\pi,\pi)$, with respect
to the crystal axes. In the $x,y$-coordinate system, ${\bm
Q}=(\pi,\pi)$ for $(100)$ interface and ${\bm Q}=(\sqrt{2}\pi,0)$
in the $(110)$ case. The quasiparticle velocity $v_{a,b}=2t\sin
k_{a,b}$ changes its sign when the momentum varies by ${\bm Q}$.
Hence, quasiparticles experience retroreflection at AF/N
interfaces. Since $v_{F,y}=0$ for a (110) interface at
half-filling, in this particular case quasiparticles move along
the interface normal, so that retroreflection and specular
reflection coincide. We note that, beyond the quasiclassical
approximation, $\bm Q$ is not a difference between incoming and
outgoing momenta anymore and the condition for retroreflection is
broken.

For energies below the quasiparticle gap in the antiferromagnet we
find the following reflection amplitudes for electrons at the AF/N
interface (for both orientations):
\begin{equation}
r_{AF,\sigma} = \frac{(\mu+\varepsilon)-i\sqrt{m^2-(\mu +
\varepsilon)^2}}{\sigma m} , \quad |\mu+\varepsilon|<m .
\label{r}
\end{equation}
Reflection amplitudes for quasiparticle energies above the gap
$|\mu+\varepsilon|>m$ are obtained from Eq.(\ref{r}) with the substitution
$i\sqrt{m^2-(\mu + \varepsilon)^2}\to {\rm sgn}(\mu + \varepsilon)
\sqrt{(\mu + \varepsilon)^2-m^2}$.

The retroreflection at an AF/N interface is analogous, to some
extent, to Andreev reflection at an SC/N interface. In particular,
bound states arise in AF/N/AF mesoscopic systems for energies
below the antiferromagnetic gap, analogously to Andreev subgap
states in SC/N/SC systems. The eigenenergies of these states are
determined by the equation $\varepsilon+\mu =
\frac{|v_{F,x}|}{d}(\varphi + \pi n)$, with integer $n$ and
$\varphi= \arccos[({\varepsilon + \mu})/{m}]$. For sufficiently
large width $d\gg\xi_m$ of the normal metal layer, and considering
quasiparticles with subgap energies $|\varepsilon+\mu|\ll m$, we
obtain the simple,  explicit  expression $\varepsilon_n + \mu
=\frac{\pi |v_{F,x}|}{d}\, (n+\frac{1}{2})$. As is known, in
SC/N/SC systems electrons and holes with the same momenta and
opposite velocities form Andreev bound states, which carry
electric current. However, due to different nature of AF and SC
order parameters, in AF/N/AF systems electrons with momenta $\bm
k$ and ${\bm k}+ {\bm Q}$ and opposite velocities are coherently
entangled and the bound states themselves do not carry electric
current.

Until now we have not discussed nonmagnetic channels for
quasiparticle reflection, which are present both on account of
Fermi velocity mismatch in a normal metal and an antiferromagnet
and/or due to potential barriers at the interface.
Potential barriers themselves form specular quasiparticle
reflection, as usual. If the characteristic parameters $t$ and
$\mu$ in two identically oriented half-spaces are close to each
other, the reflection coefficient of conventional specular
reflection arising due to a mismatch of Fermi velocities is of
order  $(m/t)^2$. Furthermore, crystal momentum can change in a
reflection process by a reciprocal crystal vector along the
surface. Due to a difference between reciprocal crystal vectors at
the interface and in the bulk, specific crystal periodicity along
a particularly oriented surface or interface can result in
additional channels for quasiparticle reflection
\cite{gantmakher}.

{\it AF/SC interface.} Eqs.(\ref{andr}) apply also to AF/SC
interfaces. The solution of these equations in the absence of
potential barriers and/or a Fermi velocity mismatch shows
zero-energy interface states in the system. These states take
place for arbitrary relation between $m\ll \varepsilon_F$ and
$\Delta_s\ll \varepsilon_F$ and for $|\mu|<m$, which garantees the
existence of the antiferromagnetic gap for electrons and holes.
The zero-energy states at the AF/sSC interface arise as a combined
effect of Andreev reflection from the superconducting halfspace
and the antiferromagnetic retroreflection from the
antiferromagnetic side. The origin of the zero-energy surface
states is closely connected with the magnetic properties of the
quasiparticle reflection. This can be easily seen in the
particular case $m\gg \Delta_s$, where quasiparticle reflection
from the antiferromagnet can be described with reflection
amplitudes (\ref{r}) for the AF/N interface. According to
Eq.(\ref{r}), antiferromagnetic ordering results in opposite signs
of reflection ampltudes for electrons with spin up and down.
Quasiparticles with energies below the antiferromagnetic gap do
not penetrate in the bulk of the itinerant AF,
$r_{AF,\sigma}=\exp(i\Theta_\sigma)$. This takes place in the
absence of potential barriers, i.e. for a transparent interface.
Quasiparticles in the superconducting halfspace can be described
in terms of standard Andreev equations for Andreev amplitudes
$\tilde{\psi}^{T}_i({\bm k}_F)\equiv (\tilde{u}_{i\uparrow}({\bm
k}_F),\tilde u_{i\downarrow}( {\bm
k}_F),\tilde{v}_{i\uparrow}({\bm
k}_F),\tilde{v}_{i\downarrow}({\bm k}_F))$ complemented with the
boundary conditions
${\tilde\psi}^{out}_{i=0}({k}_{y}+{Q}_y)={\check
S}\tilde{\psi}^{in}_{i=0}({k}_{y})$. The Andreev amplitudes
$\tilde{\psi}^{in}$ contain solutions for quasiparticles moving
towards the interface ($v_{F,x}<0$), in contrast with
$\tilde{\psi}^{out}$. The ${\check S}$-matrix for the AF/N
boundary takes the form ${\check
S}=\Bigl(r_{AF,\uparrow}\dfrac{1+\sigma_3}{2}+r_{AF,\downarrow}
\dfrac{1-\sigma_3}{2}\Bigr)\tau_0$. Thus, the problem of a
superconducting halfspace with AF/SC interface can be formulated
for quasiparticles below the antiferromagnetic gap in a form
identical to that obtained for an impenetrable ferromagnetic
surface. The spectrum of Andreev bound states $\varepsilon_B=
\pm\Delta\cos(\Theta/2)$, where
$\Theta=\Theta_\uparrow-\Theta_\downarrow$, is well studied for
the latter problem \cite{fog00}. As follows from Eq.(\ref{r}), for
the antiferromagnetic boundary $\Theta=\pi$. This leads, indeed,
to  zero-energy surface bound states on the AF/sSC interface.

Let now a $d$-wave superconductor make
a junction with an antiferromagnet without
potential barriers and/or substantial Fermi velocity mismatch. The
AF/dSC interface is closed for quasiparticles in the
superconductor with energies below the antiferromagnetic gap.
However, no low-energy surface bound states with
$|\varepsilon_B|\ll m$ (in particular, no zero-energy states) form
in the case in question for any interface-to-crystal orientation.
This  follows  directly from solutions of Eqs.(\ref{andr}) and
the continuity of Andreev amplitudes at the boundary.
This can be also explained qualitatively as a combined result of the
antiferromagnetic retroreflection and the change of
sign of the $d$-wave order parameter.  The
difference $\pi$ between phases of reflection amplitudes (\ref{r})
can be effectively ascribed to the variation of the phase of the
order parameter also. In order to see this, one can introduce
auxiliary quantities $\tilde{u}^{au}_\sigma(x,\tilde{\bm
k}_F,\varepsilon)= \tilde{u}_\sigma(x,\tilde{\bm
k}_F,\varepsilon) e^{-i\sigma\Theta/2}$,
$\tilde{v}^{au}_{\bar\sigma}(x,\tilde{\bm k}_F,
\varepsilon)=\tilde{v}_{\bar\sigma}(x,\tilde{ \bm
k}_F,\varepsilon) e^{i\sigma\Theta/2}$ into Andreev equations and
boundary conditions, taken for the outgoing momentum
$\tilde{\bm k}_F$. Andreev amplitudes for incoming
momentum ${\bm k}_F$ are kept unchanged. Then the
problem becomes, formally, identical to the one for magnetically
inactive impenetrable boundary and the effective order parameter
for the outgoing momenta ${\Delta}_{eff,\sigma}(x,\tilde{\bm
k}_F)=e^{-i\sigma\Theta} \Delta(x,\tilde{\bm k}_F)$.
Since the $d$-wave order parameter $\Delta_d({\bm k}_F,x_i)=
2\Delta_d^{i i+{\hat a}}(\cos k_a-\cos k_b)$  changes its sign when
the wavevector changes by ${\bm Q}$, we obtain
$\Delta_d(x,\tilde{\bm k}_F)=-\Delta_d(x,{\bm k}_F)$ for
$\tilde{\bm k}_F={\bm k}_F+{\bm Q}$.
Thus, an outgoing quasiparticle sees an effective superconducting order
parameter with an additional phase $\pi-\Theta$ as compared with the
order parameter for the incoming trajectory. For the antiferromagnetic
interface $\Theta=\pi$ and the total phase variation of the
effective order parameter in a reflection event vanishes.

Interface potential barriers open a channel of
specular reflection.  A combined description of
specular reflection and retroreflection includes more complicated
boundary conditions, where $S$-matrix connects
$\tilde{\psi}^{out}_0({ k}_y),\tilde{\psi}^{out}_0({k}_y+Q_y)$
with $\tilde{\psi}^{in}_0({ k}_y),\tilde{\psi}^{in}_0({ k}_y+Q_y)$
and contains both off-diagonal and diagonal components in momentum
space. This modifies the effects of retroreflection.
The zero-energy surface states will
be split at an AF/I/sSC interface and reach the value of the
superconducting gap in the limit of impenetrable insulating
interlayer.  The opening of conventional channels of
reflection on the AF/I/dSC interface will result in
interface subgap states. For a (110) interface orientation, the
subgap states will evolve with decreasing  barrier transparency to
the well known zero-energy surface states in $d$-wave
superconductors. For large potential barriers, the antiferromagnet
weakly splits the zero-energy states. A mismatch of Fermi
velocities similarly results in conventional channels of
reflection. For a small mismatch, the
energies of subgap states on the AF/dSC interface will be situated
close to the superconducting gap. If the mismatch (e.g. the
magnetization) increases, the subgap states on the (110) interface
will move towards lower energies. We believe this to be the origin
of the subgap states on the AF/dSC (110) interface obtained
recently in Ref.\ \onlinecite{andersen} on the basis of numerical
studies for a large mismatch of Fermi surfaces.

{\it Josephson current through SC/AF/SC junction.}
The zero-energy surface states arising on the AF/sSC interface
will be split for an AF interlayer with finite width $l$. The split
can be considered as an effect of tunneling between zero-energy
states on two boundaries of the AF layer. If no potential barriers
are present on the boundaries and $l\ll\xi_s$, $m\gg\Delta_s$, we
find the following energies for interface states:
$\varepsilon_B=\pm\sqrt{D}|\Delta_s\cos(\chi/2)|$, where
$D(k_y)=4K(k_y)(K(k_y)+1)^{-2}$ is the transparency of the N/AF/N
junction and $K(k_y)=\exp(2ml/|v_{F,x}(k_y)|)$. These states carry
the Josephson current
\begin{equation}
J=\int\limits_{-\pi/2}^{\pi/2}\dfrac{dk_y}{\pi}{2e\sqrt{D}|\Delta_s|\sin
\dfrac{\chi}{2}}\tanh\dfrac{\sqrt{D}|\Delta_s|
\cos\dfrac{\chi}{2}}{2T}\enspace ,
\label{jos}
\end{equation}
which differs from the Ambegaokar-Baratoff result. In the
particular case of large interlayer width, $K,D\ll 1$, there are
low-energy states in the junction which result in low-temperature
anomalous behavior of the critical current. This behavior is similar
to what can happen in tunnel junctions with $d$-wave superconductors
or S/F/S junctions with low-energy interface states \cite{bbr96,fog00,bb02}.

{\it Conclusions.} We have developed the quasiclassical theory of
itinerant antiferromagnets under the assumption of slow spatial
variations of the {\it sublattice} magnetization and near-nesting
of the metallic band. The quasiclassical equations, written for
pairs of entangled quasiparticle trajectories, provide a powerful
tool for study of a host of new problems related to interfaces
with antiferromagnetic materials. As important applications of
this formalism, we have shown that quasiparticles in normal metals
are retroreflected from the antiferromagnet, and studied bound
state formation at interfaces with superconductors. We have
further shown that Andreev zero-energy surface states are formed
at an interface between an antiferromagnet and an $s$-wave
superconductor.  These bound states do not arise on interfaces
with a $d$-wave superconductor for any interface orientation. We
demonstrated that the Josephson current in sSC/AF/sSC junction is
strongly influenced by the low-energy interface states and differs
from the conventional Ambegaokar-Baratoff result. In analogy with
other situations \cite{barash00}, one can expect that, after
introducing some effective order parameters, the self-consistent
calculations do not modify low-energy bound states and, in
particular, their contribution to the Josephson current. Our
formalism in the present paper is based on the Andreev equations,
but we have derived also the Eilenberger equations for itinerant
antiferromagnets and respective boundary conditions. Based on
these equations, we plan to explore with self-consistent
calculations proximity effects for AF/SC interfaces as well as the
problems of the Josephson current and the low-bias conductance in
a longer work.

{\it Acknowledgments.} We thank S.N. Artemenko and V.V. Ryazanov
for useful discussions. This work was supported by grant
NSF-INT-0340536 (I.V.B., P.J.H., and Yu.S.B.). I.V.B. and Yu.S.B.
also acknowledge the support by grant RFBR 02-02-16643
and scientific programs of Russian Ministry of Science and Education
and Russian Academy of Sciences. I.V.B. thanks the support from
Forschungszentrum J\"ulich (Landau Scholarship) and the Dynasty Foundation.

\begin{thebibliography}{99}
%
\bibitem{ryazanov01}
V.V.~Ryazanov {\it et al}.,
Phys. Rev. Lett. {\bf 86}, 2427 (2001).
%
\bibitem{aprili02}
T.~Kontos {\it et al}.,
Phys. Rev. Lett. {\bf 89}, 137007 (2002).
%
\bibitem{aprili03}
W.~Guichard {\it et al}.,
Phys. Rev. Lett. {\bf 90}, 167001 (2003).
%
\bibitem{aprili04}
A.~Bauer {\it et al}.,
Phys. Rev. Lett. {\bf 92}, 217001 (2004).
%
\bibitem{buz82}
A.I.~Buzdin {\it et al}., JETP Lett. {\bf
35}, 178 (1982) [Pis'ma~Zh.~Eksp.~Teor.~Fiz. {\bf 35}, 147
(1982)].
%
\bibitem{buz92}
A.I.~Buzdin {\it et al}., Sov.~Phys.~JETP
{\bf 74}, 124 (1992) [Zh.~Eksp.~Teor.~Fiz. {\bf 101}, 231 (1992)].
%
\bibitem{nazarov01}
N.M.~Chtchelkatchev {\it et al}.,
JETP Lett. {\bf 74}, 323 (2001) [Pis'ma Zh.~Eksp.~Teor.~Fiz. {\bf
74}, 357 (2001)].
%
\bibitem{bb02}
Yu.S.~Barash, I.V.~Bobkova,
Phys.~Rev.~B {\bf 65}, 144502 (2002).
%
\bibitem{bb01}
Yu.S.~Barash {\it et al}.,
Phys. Rev. B {\bf 66}, 140503 (2002).
%
\bibitem{mrs88}
A.~Millis {\it et al}.,
Phys. Rev. B {\bf 38}, 4504 (1988).
%
\bibitem{fog00}
M.~Fogelstr\"om, Phys. Rev. B {\bf 62}, 11812 (2000).
%
\bibitem{blamire03}
C.~Bell {\it et al}.,
Phys. Rev. B {\bf 68}, 144517 (2003).
%
\bibitem{gorkov01}
L.P.~Gor'kov, V.~Kresin,
Appl. Phys. Lett. {\bf 78}, 3657 (2001).
%
\bibitem{kasatkin84}
A. L. Kasatkin, E. A. Pashitskii, Sov. J. Low Temp. Phys. {\bf 10},
640 (1984); Sov. Phys. Solid State {\bf 27}, 1448 (1985).
%
\bibitem{visscher96}
M.I. Visscher, G.E.W. Bauer, Phys. Rev. B {\bf 54}, 2798 (1996).
%
\bibitem{rejaei96}
B. Rejaei, G.E.W. Bauer, Phys. Rev. B {\bf 54}, 8487 (1996).
%
\bibitem{ar97}
S.N. Artemenko and S.V. Remizov, Pis'ma Zh. Eksp. Teor. Fiz. {\bf 65},
50 (1997) [JETP Lett. {\bf 65}, 53 (1997)].
%
\bibitem{latyshev96}
A.A. Sinchenko {\it et al}.,
JETP Lett. {\bf 64}, 285 (1996); Phys. Rev. B {\bf 60}, 4624 (1999).
%
\bibitem{gantmakher}
V.F. Gantmakher, Y.B. Levinson, {\it Carrier Scattering in Metals and
Semiconductors}, Elsevier Science Ltd., 1987.
%
\bibitem{andersen}
B.M.~Andersen, P.~Hedegard, Phys. Rev. B {\bf 66}, 104515
(2002).
%
\bibitem{bbr96}
Yu.S.~Barash {\it et al}.,
Phys.~Rev.~Lett. {\bf 77}, 4070 (1996).
%
\bibitem{barash00}
Yu.S.~Barash,
Phys. Rev. B {\bf 61}, 678 (2000).
%
\end{thebibliography}

\end{document}